\documentclass[conference]{IEEEtran}
\usepackage{amsmath,amssymb,eucal,graphicx}
\usepackage{pst-plot}
\usepackage{epsfig}
\usepackage{amsthm}
\usepackage{amsfonts}
\usepackage{epsfig}
\usepackage{float}
\usepackage{pstricks}
\usepackage{color}
\usepackage{multicol,subfigure}
\usepackage{mathcomSTEP,stmaryrd}

\theoremstyle{remark}
\newtheorem{rem}[thm]{Remark}

\begin{document}

\title{%
On the Capacity of a General Multiple-Access Channel and of a Cognitive Network in the Very Strong Interference Regime
}
\author{%
\IEEEauthorblockN{Stefano Rini}
\IEEEauthorblockA{
Lehrstuhl f\"{u}r Nachrichtentechnik,  Technische Universit\"{a}t M\"{u}nchen \\
Arcisstra{\ss}e 21, 80333 M\"{u}nchen, Germany,  Email: \tt stefano.rini@tum.de}
}

\maketitle

\begin{abstract}

The capacity of the multiple-access channel with any distribution of messages among the transmitting nodes was determined by Han in 1979
and the expression of the capacity region contains a number of rate bounds and that grows exponentially with the number of messages.
We derive a more compact expression for the capacity region of this channel
in which the number of rate bounds depends on the distribution of the messages at the encoders.
%
%
%
%
Using this expression we prove capacity for a class of general cognitive network that we denote as ``very strong interference'' regime.
%
%
%
In this regime there is no rate loss in having all the receivers decode all the messages
and the capacity region reduces to the capacity of the compound multiple-access channel.
This result generalizes the ``very strong interference'' capacity results for the interference channel, the cognitive interference channel, the interference channel with  a cognitive relay and many others.

\end{abstract}

{\IEEEkeywords
multiple access channel, cognitive network, strong interference, superposition coding, cut-set bound.
}

\section{Introduction}

The multiple access channel is a well studied channel model where multiple transmitters communicate with a single receiver.
The capacity region of the two-users multiple-access channel
was characterize in 1971  by Ahlswede  \cite{ahlswede1971multi} and Liao \cite{liao1972multiple}.
%
Slepian and Wolf \cite{slepian1973coding} studied a more general two-users channel  where the transmitters, in addition to their own message, also cooperate in communicating a common message.
%
%
Han \cite{han1979capacity} generalized all the previously known results and derived the capacity region of a general multiple-access channel with any distribution of the messages at the encoders.
%
%
The  expression of the capacity region contains an auxiliary random variables for each of the messages to be transmitted and a number of rate bounds that grows exponentially with the number of messages.
Despite of this result, it is known that, in certain special cases,  it is possible to describe the capacity region with less rate bounds and fewer random variables.
Two such examples are the multiple-access channel with independent messages \cite{ahlswede1971multi,liao1972multiple} and
the degraded multiple-access channel \cite{HaroutunianLowerBoundMAC}.
G\"{u}nd\"{u}z and Simeone \cite{gunduz2010capacity}
derived a compact expression for the capacity region of a multiple-access channel with a specific distribution of the messages among the transmitters that
encompasses the examples above.
%
The authors also describe an algorithmic procedure to convert a general multiple access channel in an multiple-access channel with the specific message distribution they consider, but provide no closed form expression for the general case.

%
%
%
%

One can sometimes exploit the available capacity results  for the general multiple-access channel to derive the capacity for a subclass
of general channels in what is generally referred to as the ``very strong interference'' regime.
%
%
This regime is characterized by the fact that the level of the interference at each decoder is so high to allow for the decoding of all the interfering signals and the complete cancelation of the interference
%
%
As a result, the capacity region of the channel reduces to the intersection of the multiple-access channels between the transmitters and each receiver.
%
%
%
The first ``very strong interference'' result was derived for the two user interference channel Carleial \cite{carleial1975case}  and Sato \cite{sato.IFC.strong}.
Inspired by this result, similar ``very strong interference'' capacity results was proved for the two user cognitive interference channel \cite{maric2005capacity}, the interference channel with a cognitive relay \cite{rini2011capacity} and others.
An interesting ``very strong interference'' result is provided by Sridharan et al. for  the Gaussian symmetric K-user interference channel \cite{sridharan2008capacity} by employing lattice codes: here the receivers do not decode each interfering signal separately, but instead decode their total sum.
%
We note that this result heavily relies on the structure Gaussian channel and does not extend to a general channel model.

\noindent
{\bf Contributions and Paper Organization :}

\noindent
{\bf Sec. \ref{sec:Network Model}--We introduce a  general cognitive network} and the notation used throughout the paper.

%

\noindent
{\bf Sec. \ref{sec:Inner and Outer Bounds for a General Cognitive Network}--We provide inner and outer bounds for the general cognitive network}. The outer bounds is reminiscent of the cut-set outer bound and the inner bound is based on rate splitting and superposition coding

\noindent
{\bf Sec. \ref{sec:Capacity for a general Multi-Access Channel}--We provide a compact characterization of the capacity for a general multiple-access channel } which requires less rate bounds than \cite{han1979capacity} and is valid for a general distribution of messages, unlike \cite{gunduz2010capacity}.

\noindent
{\bf Sec. \ref{sec:Capacity for a general cognitive network}--We show capacity for a general cognitive network in ``very strong interference'' regime} 
in which capacity is achieved employing superposition coding and having all the decoders decode all the messages.

\noindent
{\bf Sec. \ref{sec:An example: the interference channel with common messages}--We provide an example of our results} by deriving capacity in the ``very strong interference'' regime for the interference channel where each transmitter is sending a private and a common message.

\section{Network Model}
\label{sec:Network Model}

We consider the general cognitive multiple-terminal network in \cite{rini2011achievable} where
$N_{\rm TX}$ transmitting nodes want to communicate with $N_{\rm RX}$ receiving nodes.
A given node may only be a transmitting or a receiving node, that is,
the network is single-hop and without feedback or cooperation.
The transmitting node $k$, $k \in [1: N_{\rm TX}]$, inputs $X_k$ to the channel, while the receiving node $z$, $z \in [1 : N_{\rm RX}]$,
has access to the channel output $Y_z$.
The channel transition probability is indicated with $P_{Y_1  :  Y_{N_{\rm RX}}|X_1  :      X_{N_{\rm TX}}}$ and
the channel is assumed to be memoryless.
The subset of transmitting nodes $\iv$, $\iv \subset \PS_{N_{\rm TX}}$, is interested in sending the message
$W_{\iv \sgoes \jv}$ to the subset of receiving nodes $\jv \subset \PS_{N_{\rm RX} }$ over $N$ channel uses.
%
%
The message $W_{\iv \sgoes \jv}$, $(\iv,\jv) \subset \PS_{N_{\rm TX}} \times \PS_{N_{\rm RX} }$, uniformly distributed Random Variable (RV)  in the interval  $[0  : 2^{N R_{\iv \sgoes \jv}}-1]$, where $N$ is the block-length and $R_{\iv \sgoes \jv}$ the transmission rate.

A rate vector $\Rv= \{ R_{\iv \sgoes \jv}, \ \forall \ (\iv,\jv) \subset   \PS_{N_{\rm TX}} \times \PS_{N_{\rm RX}}\}$ is said to be achievable if there exists
a sequence of encoding functions
\pp{
X_k^N = X_k^N,
\lb \lcb
W_{\iv \sgoes \jv}, \ST   \ R_{\iv \sgoes \jv} \in \Rv,  k \in \iv \rcb \rb,
}
and a sequence of decoding  functions
\ea{
\widehat{W}_{\iv \sgoes \jv}^z =\widehat{W}_{\iv \sgoes \jv}^z \Big(Y_z^N\Big)\ \text{if}\ z \in \jv,
}
such that
\begin{align*}
\lim_{N \to \infty } \max_{\iv,\jv,z} \Pr\lsb  \widehat{W}_{\iv \sgoes \jv}^z \neq W_{\iv \sgoes \jv}^z  \rsb = 0.
\end{align*}
The capacity region $\Cc(\Rv)$ is the convex closure of the region of all achievable rates in the vector $\Rv$-pairs.

\begin{figure}[h!]
\centering
\vspace{- .5 cm}
\includegraphics[width=9.5 cm]{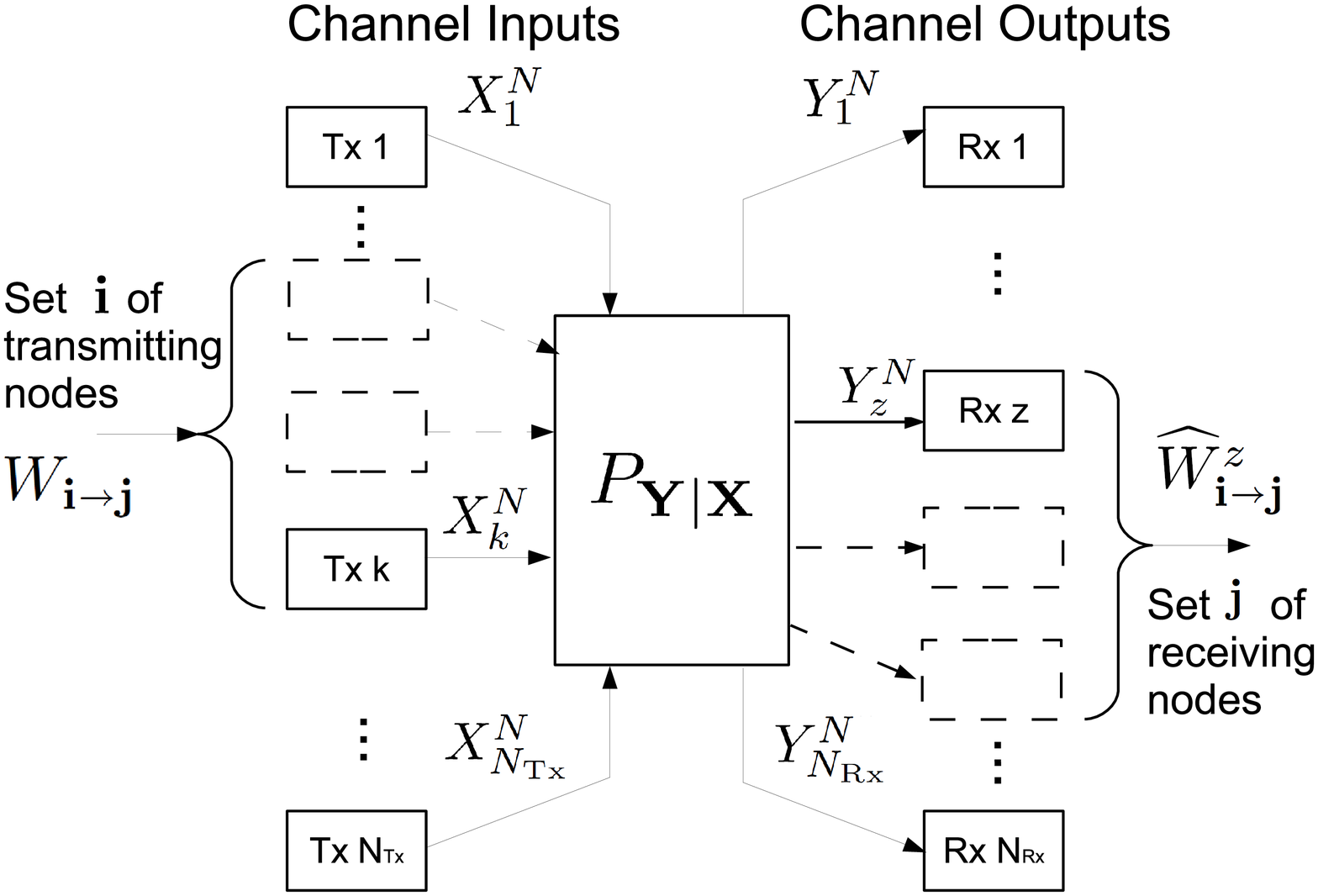}
\vspace{- 1.4 cm}
\caption{The general cognitive network.}
\label{fig:channelModel}
\vspace{- .5 cm}
\end{figure}

The general network model we consider is a variation to the network model in \cite[Ch. 14]{cover1991elements} but we allow for messages ho be distributed to more
than one user while disregarding feedback.
%
Fig. \ref{fig:channelModel} depicts a general cognitive network.
%

In the following we focus in particular to the general Multiple-Access Channel (MAC), that is, the channel where there is only one receiver ($N_{\rm Rx}=1$).

\subsection{Notation}
\label{sec:Notation}

When considering the capacity problem $\Ccal\lb \Rv_{\iv \sgoes \jv}\rb$ we use the following notation
\begin{itemize}
  \item   $\Sv$ is a set containing $(\iv,\jv)$ pairs:
\ea{
\Sv \subset \lcb (\iv,\jv) \ \ST \ R_{\iv \sgoes \jv} \in \Rv \rcb,
\label{eq:Sv definintion}
}
and $\Svo$ indicates the complement of the set.
  \item $\Tv_{\Sv}$  is the set of transmitters that can conveys \emph{all} the messages in $\Sv$:
\ea{
\Txv_{\Sv}=\lcb k ,  k \in  \bigcap_{  (\iv,\jv) \in \Sv } \iv      \rcb,
}

\item $\Rv_{\Sv}$ is the set of all the receivers that decode \emph{any} of the messages in $\Sv$
\ea{
\Rxv_{\Sv}=\lcb z ,  z \in  \bigcup_{  (\iv,\jv) \in \Sv } \jv      \rcb.
}

\end{itemize}

\section{Inner and Outer Bounds for a General Cognitive Network}
\label{sec:Inner and Outer Bounds for a General Cognitive Network}

%

We begin by stating outer bounds for the general cognitive network model in Sec. \ref{sec:Network Model} that was originally devised in \cite[Th.  5.1]{han1979capacity}.
This outer bound generalizes the outer bound in \cite[Th 3.2]{gunduz2010capacity} to any number of users
and is reminiscent of the cut-set bound in \cite[Th. 14.10.1]{cover1991elements}
when extended to the case where messages are distributed among the users.

\begin{cor} {\bf Cut-Set Bound for a General Cognitive Network}
\label{cor:min cut outer bound}
If the rate vector $\Rv$ is achievable for the general multi-terminal network in Sec. \ref{sec:Network Model},  the following must hold
\ea{
&0 \leq  \sum_{(\iv,\jv) \in \Sv} R_{\iv \sgoes \jv} \leq \sum_{z \in \Rxv_{\Sv}} I \lb Y_z ; \lcb X_k, \ \forall \ k  \rcb  \rnone
\label{eq:th:min cut outer bound with channel inputs with auxiliary RVs} \\
&  \quad \quad  \quad \quad \quad \lnone  | \lcb X_k, \ k \in \Txv_{\Svo^z}    \rcb , \lcb U_{\iv \sgoes \jv}, \ (\iv, \jv)\in \Svo^z \rcb \rb, \nonumber
}
for every $\Sv$ as defined in \eqref{eq:Sv definintion} and any partition $\{\Sv^z\}$ of $\Sv$ with
\ea{
(\iv,\jv) \in \Sv^z  \implies  z \in \jv,
\label{eq:partitioning Sv in the general outer bound with channel inputs}
}
and any distribution
\ea{
&P_{\lcb X_k, \ \forall \ k  \rcb ,  \lcb U_{\iv \sgoes \jv}, \ \forall (\iv,\jv) \rcb }=
\prod_{ (\iv,\jv)}P_{U_{\iv \sgoes \jv} }  \prod_{k} P_{X_k | \{ U_{ \iv \sgoes \jv}, \ k \ \in \iv \} }.
\label{eq:factorization auxiliary U outer bound}
}
\end{cor}
\begin{IEEEproof}
See \cite{han1979capacity} for a complete proof.
In essence Fano's inequality is applied to any set of messages $W_{\iv \sgoes \jv}, \ (\iv,\jv) \in \Sv$ as
\eas{
 \sum_{(\iv,\jv) \in \Sv}& R_{\iv \sgoes \jv}   \leq  \sum_{z \in \Rxv_{\Sv}} I \lb Y_z^N  ; \{W_{\iv \sgoes \jv}\in \Sv^z \} \rb \\
& \leq  \sum_{z \in \Rxv_{\Sv}} I \lb Y_z  ; \{W_{\iv \sgoes \jv}\in \Sv^z \} | \{W_{\iv \sgoes \jv}\in \Svo^z \} \rb,
}
where $\Sv^z$ determines which $W_{\iv \sgoes \jv}$ are decoded at receiver $z$ (note that the complement of $\Sv$ is taken with respect to all the messages in the network).
The single letterization follows from the memoryless property of the channel and the bound in \eqref{eq:partitioning Sv in the general outer bound with channel inputs} is obtained defining the auxiliary RV $U_{\iv \sgoes \jv}=W_{\iv \sgoes \jv}$.
\end{IEEEproof}
%
%
%
We now derive  a general achievable region  obtained with the chain graph representation of an achievable scheme in \cite{rini2011achievable}
\begin{cor}{\bf An Achievable Region Based on Superposition Coding and Rate Splitting}
\label{cor:Superposition coding achievable scheme}
A rate vector $\Rv$ is achievable in the general network in Sec. \ref{sec:Network Model} if there exists a rate vector $\Rv_{\rm all}$ such that
\ea{
\jv_{\rm all} & = \bigcup_{\jv \in \Rv} \jv \quad \quad
\Rv_{\rm all}   = \lcb R_{\iv \sgoes \jv_{\rm all}}, \ \iv \in \Rv \rcb \nonumber \\
R_{\iv \sgoes \jv_{\rm all}} & = \sum_{ \jv } R_{\iv \sgoes \jv},
\label{eq:rate splitting all common}
}
that satisfies
\ea{
\sum_{(\iv,\jv_{\rm all}) \in \Sv' } R_{\iv \sgoes \jv_{\rm all}} \leq
I \lb Y_z ; \lcb X_k, \ \forall \ k\rcb  |  \lcb U_{\iv \sgoes \jv_{\rm all}}',  (\iv, \jv_{\rm all} \in \overline{\Sv'})\rcb \rb,
\label{eq:Superposition coding achievable scheme}
}
for all the subsets $\Sv' \subset \lcb (\iv, \jv_{\rm all}), \ (\iv,\jv_{\rm all})\in \Rv_{\rm all} \rcb$ such that
\ea{
(\iv,\jv_{\rm all}) \in \Sv' \implies (\lv,\jv_{\rm all}) \in \Sv', \  \forall  \ \lv \subset \iv,
\label{eq:condition joint decoding}
}
for all decoders $z$  and for any distribution
\ea{
& P_{ \{X_k, \ \forall \ k \},\{U_{\iv \sgoes \jv^{\rm all}}',  \ (\iv , \jv_{\rm all}) \in \Rv_{\rm all}  \}} = \nonumber \\
& \quad \prod_{(\iv,\jv_{\rm all})} P_{U_{ \iv \sgoes \jv_{\rm all}}' | \lcb U_{ \lv \sgoes \jv_{\rm all}}', \ \lv \supset \iv \rcb }
\cdot \prod_{k} P_{X_k | \{ U_{ \iv \sgoes \jv_{\rm all}}', \ k \in \iv \}}.
\label{eq:factorization auxiliary U inner bound}
}
\end{cor}

\begin{IEEEproof}
The theorem is a special case of  the general achievable region in \cite{rini2011achievable}.
All the messages $W_{\iv \sgoes \jv}$ are decoded at all receivers. The messages transmitted by the same set of encoders $\iv$ are encoded in the codeword
$U_{\iv \sgoes \jv}'^N$ with rate $R_{\iv \sgoes \jv_{\rm all}}$, where subscript $|_{\rm all}$ indicates that the codeword is decoded by all receivers.
Equation \eqref{eq:rate splitting all common} describe the rate splitting strategy where all the messages are decoded at all the receiver.
For this reason, the messages known at the same set of encoders $\iv$ can be encoded in the same codeword $U_{\iv \sgoes \jv_{\rm all}}'^N$ with rate
$R_{\iv \sgoes \jv_{\rm all}}$.
%
%
The two codewords  $U_{\iv \sgoes \jv_{\rm all}}'^N$ are then superposed the one on top of the other if the bottom codeword is known at a larger set of encoders than the top codeword.
%
%
Intuitively, each rate bounds corresponds with the event that the all the codewords $U_{\iv \sgoes \jv^{\rm all}}'^N$ in $\Sv'$
in \eqref{eq:Superposition coding achievable scheme} has been incorrectly decoded while the codewords in $\Svo'$ are correctly decoded.
Not all the possible decoding errors are possible, though.
When a codeword  $U_{\iv \sgoes \jv^{\rm all}}'^N$ has been incorrectly decoded, all the codewords superposed to $U_{\iv \sgoes \jv^{\rm all}}'^N$ are incorrectly decoded as well. This condition is expressed by equation \eqref{eq:condition joint decoding}.
\end{IEEEproof}

\section{A New Formulation of Capacity for a General Multiple-Access Channel}
\label{sec:Capacity for a general Multi-Access Channel}

We begin by stating the formulation of the capacity region of a general MAC derived in \cite[Th.  5.1]{han1979capacity}

\begin{cor}{\bf The Capacity Region of a General Multiple-Access Channel \cite[Th.  5.1]{han1979capacity}}
\label{cor:Capacity for General Multiple Access Channel Han}
The capacity of the general MAC is
\ea{
\sum_{ (\iv, z ) \in \ \Sv} R_{\iv \goes z } \leq
I \lb Y_{z }; \lcb X_k, \ \forall \ k \rcb  | \lcb  U_{\iv \sgoes z} , \ (\iv,z) \in \Svo \rcb  \rb,
\label{eq:Capacity for General Multiple Access Channel}
}
for all  the sets $\Sv$ in \eqref{eq:Sv definintion} and taken the union over all the distributions in
\eqref{eq:factorization auxiliary U outer bound}.
\end{cor}

\begin{thm}{\bf Compact Formulation of the Capacity Region of the Multiple Access Channel}
\label{th:Capacity for General Multiple Access Channel}
The capacity of the general MAC is given by \eqref{eq:Capacity for General Multiple Access Channel}
for all the sets $\Sv$ such that \eqref{eq:condition joint decoding} for $\Sv=\Sv'$ holds and taken the union over all the distributions in
\eqref{eq:factorization auxiliary U inner bound} for $U_{\iv \sgoes z}=U_{\iv \sgoes \jv_{\rm all}}'$.
\end{thm}
\begin{IEEEproof}
We show capacity by matching each of the rate bounds in the inner bound expression in \eqref{eq:Superposition coding achievable scheme} with a outer bound expression in \eqref{eq:th:min cut outer bound with channel inputs with auxiliary RVs} and then consider the union over all the possible distributions $P_{U_{\iv \sgoes \jv}}$ in the two regions.
%
%
%
Since there is only decoder in the network, $z$, only one permutation $\{ \Sv^z \}$  is possible and it is $\Sv^z=\Sv$.
Again, given that there is only one decoder, we have that $\j_{\rm all}=z$ and thus we can match each $\Sv$ in Cor. \ref{cor:Superposition coding achievable scheme} with the  set $\Sv^z=\Sv$ in Cor. \ref{cor:min cut outer bound} and $U_{\iv \sgoes z}=U_{\iv \sgoes \jv_{\rm all}}'$.
This shows that for each rate bound in Cor. \ref{cor:Superposition coding achievable scheme}, there exists a matching outer bound in Cor. \ref{cor:min cut outer bound}.
The proof concludes by noting that the distribution of the auxiliary RVs $U_{\iv \sgoes \jv}$ in the inner bound, \eqref{eq:factorization auxiliary U inner bound} has a more general form that the distribution in the outer bound, \eqref{eq:factorization auxiliary U outer bound}.
%
%
\end{IEEEproof}
The result in  Th. \ref{th:Capacity for General Multiple Access Channel} is a extension of \cite[Th 3.2]{gunduz2010capacity}  to the general MAC channel.
Note that the number of bounds in the formulation of the capacity region in Cor. \eqref{cor:Capacity for General Multiple Access Channel Han} grows exponentially with the number of messages since the possible sets $\Sv$ are obtained from all the permutations in \eqref{eq:Sv definintion}.
On the other hand the expression of the capacity region in Th. \ref{th:Capacity for General Multiple Access Channel} grow, in general, much less since one needs to consider only the permutations of $\Sv$ for which \eqref{eq:condition joint decoding} holds.

\begin{rem}
It interesting to compare the region in Cor. \ref{cor:Capacity for General Multiple Access Channel Han} with the region in Th. \ref{th:Capacity for General Multiple Access Channel}: in the of region Cor. \ref{cor:Capacity for General Multiple Access Channel Han} the number of bounds increases exponentially with the number of messages but the auxiliary RVs $U_{\iv \sgoes z}$ are  independent.
For the region in  Th. \ref{th:Capacity for General Multiple Access Channel} is more compact but the auxiliary RVs $U_{\iv \sgoes z}$  are no longer independent. %
It appears that one can trade a simpler expression in the capacity region at the cost of using correlated RVs in the expression of this region.
%
\end{rem}

\section{Capacity for a General Cognitive Network}
\label{sec:Capacity for a general cognitive network}
In this section we utilize the expression of capacity of the general MAC channel in Th. \ref{th: Very Strong Interference capacity results}
to derive the capacity of the general network in the ``very strong interference'' regime,
where there is no loss of optimality in having all the receivers decodes all messages.
In this class of channels the inner expression in Cor. \ref{cor:Simplified inner bound} and outer bound expression of Cor. \ref{cor:min cut outer bound} can both be simplified and shown to be equivalent.

We begin by deriving the conditions under which the outer bound expression can be simplified by replacing the bounds in  \eqref{eq:th:min cut outer bound with channel inputs with auxiliary RVs}. The outer bound obtained in this case is sometimes referred to as ``strong interference'' outer bound.

\begin{cor}{\bf Strong Interference Outer Bound}
\label{cor:strong inteference outer bound}

Consider a set $\Sv$ in Cor. \ref{cor:min cut outer bound} and a partition  $\{\Sv^z\}$, if
\ea{
& \sum_{z \in \Rxv_{\Sv}}
 I(Y_z; \lcb X_k, \ \forall \ k \rcb | \lcb U_{\iv \sgoes \jv}, \ (\iv, \jv) \in  \Svo^z \rcb) \nonumber \\
& \quad \quad \leq  I \lb Y_{z'}; \lcb X_k, \ \forall \ k \rcb  | \{U_{\iv\sgoes \jv}, \  (\iv,\jv) \in \Svo \} \rb,
 \label{eq:strong interfernce condition}
}
for all the distributions  of $\lcb X_k, \ \forall \ k \rcb$ and  $\lcb U_{\iv \sgoes \jv}, \ (\iv,\jv) \in \Rbb \rcb$
 in \eqref{eq:factorization auxiliary U outer bound}
 and for some $z'$, then the bound in \eqref{eq:th:min cut outer bound with channel inputs with auxiliary RVs} for $\Sv$, $\{\Sv^z \}$ and $z$
can be eliminated from the outer bound while the following bound is introduced
\ea{
&\sum_{(\iv,\jv) \in \Sv} R_{\iv \sgoes \jv} \geq  \\
& \quad \quad  I(Y_{z'}; \lcb X_k, \ \forall \ k \rcb  | \{U_{\iv\sgoes \jv}, \  (\iv,\jv) \in \Svo^z \}).
\label{eq:min cut outer bound with channel inputs with auxiliary RVs}
}
\end{cor}
\begin{IEEEproof}
The complete proof can found in \cite{rini2012GeneralMAC}.
\end{IEEEproof}

Note that the bound in \eqref{eq:min cut outer bound with channel inputs with auxiliary RVs} needs not be a bound of the outer bound region in Cor. \ref{cor:min cut outer bound}.
%

\begin{cor}{\bf Simplified Inner Bound}
\label{cor:Simplified inner bound}
Consider a set $\Sv'$ in Cor. \ref{cor:Superposition coding achievable scheme}, if
\ea{
& I \lb Y_z ; \lcb X_k, \ \forall \ k   \rcb  |  \lcb U_{\iv \sgoes \jv_{\rm all}}',  (\iv, \jv_{\rm all} \in \Svo')\rcb \rb \nonumber \\
&\quad   \geq \sum_{\{ \widetilde{\Sv'} , \widetilde{Y_z}\} }  I \lb \widetilde{Y_{z}} ; \lcb X_k, \ \forall k   \rcb  | \rnone  \nonumber \\
&\quad \quad \quad \quad  \quad  \lnone \lcb U_{\iv \sgoes \jv_{\rm all}}',  (\iv, \jv_{\rm all} \in \overline{\widetilde{\Sv'}})\rcb \rb,
\label{eq:Simplified inner bound}
}
for some set  $\{ \widetilde{\Sv}' \} \supset \Sv'$ such that \eqref{eq:condition joint decoding} hold for every $\widetilde{\Sv}'$ and for all the distributions  of $\lcb U_{\iv \sgoes \jv}', \ (\iv,\jv) \in \Rv \rcb$
 in \eqref{eq:factorization auxiliary U inner bound}  and for some  set of output $\{ \widetilde{Y_z}\}$,
then the bound in \eqref{eq:Superposition coding achievable scheme} for $\Sv'$ and $z$ can be eliminated from the achievable region.
\end{cor}
\begin{IEEEproof}
The complete proof can found in \cite{rini2012GeneralMAC}: the theorem states the conditions under which a rate bound in the inner bound in Th. \ref{cor:Simplified inner bound} can be eliminated
because it is larger than a linear combination of other rate bounds in the achievable region.
The each set $ \widetilde{\Sv}'$ corresponds to a rate bound in \eqref{eq:Superposition coding achievable scheme} for a channel output $\widetilde{Y_{z}}$.
The collection of sets  $ \{\widetilde{\Sv}' \}$ corresponds then to the summation of different bounds that is larger than rate bound in \eqref{eq:Simplified inner bound} for $\Sv'$ and $Y_z$.
\end{IEEEproof}

By combining the results in Cor. \ref{cor:strong inteference outer bound} and Cor. \ref{cor:Simplified inner bound} we can finally prove capacity for a general cognitive network in the  ``very strong interference'' regime.

\begin{thm}{\bf Very Strong Interference Capacity Results}
\label{th: Very Strong Interference capacity results}

Consider the achievable region in Cor. \ref{cor:Superposition coding achievable scheme} with the assignment
\ea{
U_{\iv \sgoes \jv_{\rm all}}'   = \lcb U_{\iv \sgoes \jv}, \ \forall \ (\iv,\jv) \in \Rv \rcb,
\label{eq:assigment Uprime to U}
}
so that the set $\Sv'$ in Cor. \ref{cor:Superposition coding achievable scheme}  coincides with the set $\Sv$ for
\ea{
\Sv=\lcb (\iv,\jv) , \ \iv \in \Sv', \ \jv \in \Rv  \rcb,
\label{eq:realte S and Sall}
}
and the rate vector $\Rv$ is obtained from the rate vector $\Rv_{\rm all}$ with \eqref{eq:rate splitting all common}.
This region is capacity if, for each $\Sv$ in \eqref{eq:realte S and Sall} and each $z$, one of the following conditions holds


\noindent
{\bf i)}
 $z \in \bigcap_{(\iv,\jv) \in \Sv} \jv$ so that we can set $\{ \Sv^z \}=\Sv$,

 \noindent
{\bf ii)}
    there exists a partition $\{ \Sv^z \}$ and some $z'$ for which condition \eqref{eq:strong interfernce condition}  holds, or

\noindent
   {\bf iii)}
   there exists a set $\{\widetilde{\Sv_{\rm all}}, \widetilde{Y_z}\}$ for which condition \eqref{eq:Simplified inner bound} holds.
\end{thm}

\begin{IEEEproof}
The complete proof can found in \cite{rini2012GeneralMAC}.
%
%
The expressions in  \eqref{eq:assigment Uprime to U}  and \eqref{eq:realte S and Sall} relate the auxiliary RV in the inner bound to the auxiliary RVs in the outer bound.
As also pointed out in Cor. \ref{cor:Superposition coding achievable scheme},the inner bound is produced by having all the decoders decode all the messages:  this means that the messages transmitted by the same set of encoders $\iv$ but destined to different sets of decoders $\jv$ can be embedded in a single codeword, $U_{\iv \sgoes \jv_{\rm all}}^N$.
The set $\Sv'$ describe all the possible error events when decoding $U_{\iv \sgoes \jv_{\rm all}}^N$: since this codeword encodes a multiple messages for different set of decoders$\jv$, an error in decoding $U_{\iv \sgoes \jv_{\rm all}}^N$ corresponds to an error in decoding all the messages it encodes.
This is formally expressed by \eqref{eq:realte S and Sall}: $\Sv$ is obtained from $\Sv'$ by considering all the messages $W_{\iv \sgoes \jv}$ that are encoded in $U_{\iv \sgoes \jv_{\rm all}}^N$ for a given $(\iv,\jv_{\rm all}) \in \Sv'$.
%
%
Having specified the relationship between auxiliary RVs in the inner and outer bound specified by \eqref{eq:assigment Uprime to U}  and \eqref{eq:realte S and Sall}, we can now proceed in matching inner and outer bound.  This can be done in three ways: either {\bf i)} there exists an outer bound expression matching the inner bound , or
{\bf ii)} we set the conditions to impose a matching outer bound expression
{\bf iii)} we set the condition for the inner bound expression to be redundant.
%
%
\end{IEEEproof}

\begin{rem}
From Th. \ref{th: Very Strong Interference capacity results} one concludes that capacity can be determined by imposing different conditions on $Y_z$ and $U_{\iv \sgoes \jv}$'s.
Not all the choices will be feasible though and some choices will be unfeasible for all channels but some degenerate channel.
%
\end{rem}

\section{An Example: the Interference Channel with Common Messages}
\label{sec:An example: the interference channel with common messages}

\begin{figure}[h!]
\centering
\vspace{- .75 cm}
\includegraphics[width=9.5 cm]{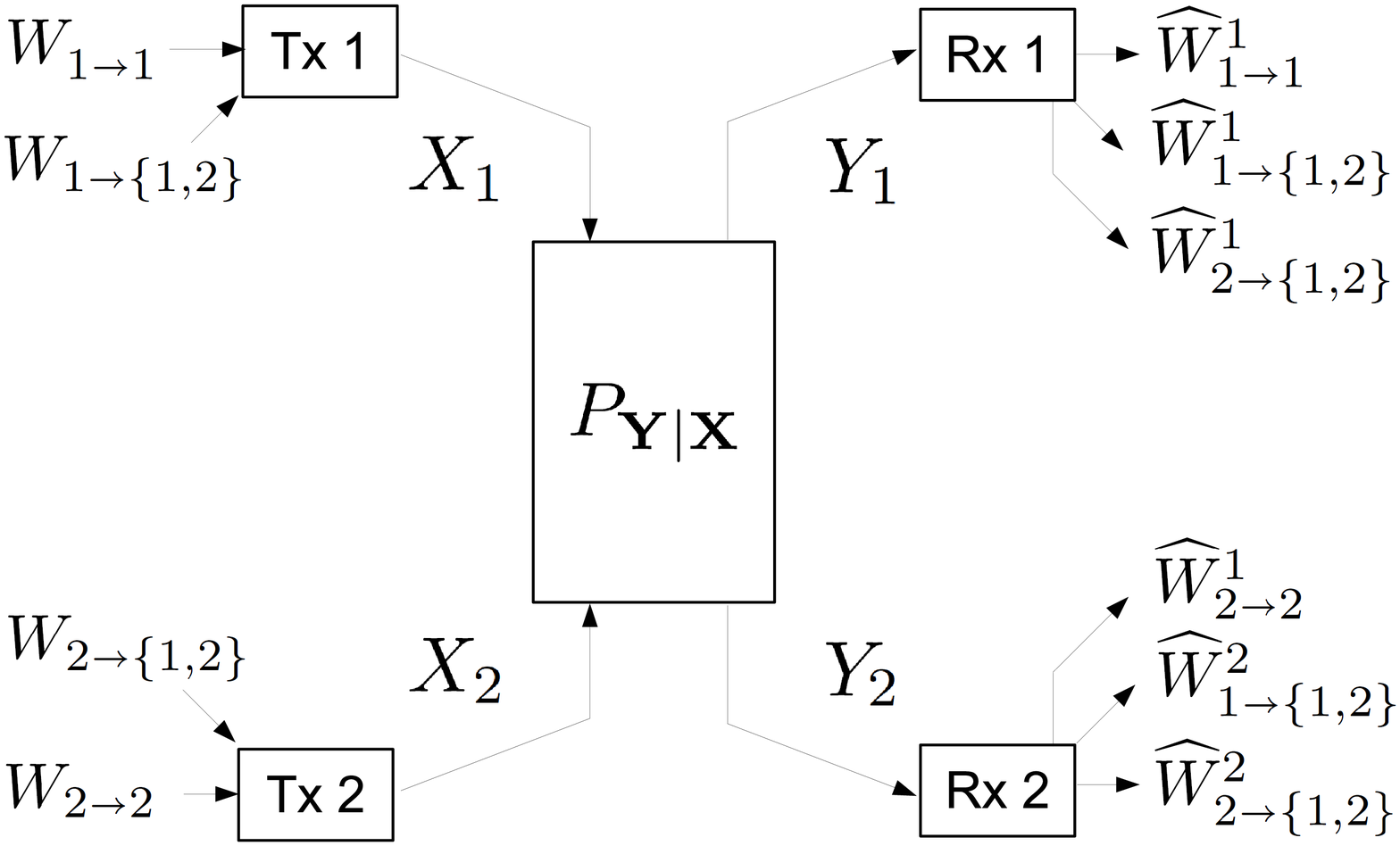}
\vspace{- 1.5 cm}
\caption{The InterFerence Channel with 2 Common Messages (IFC-2CM)}
\label{fig:IFC-CM}
\vspace{-.25 cm}
\end{figure}

We now apply the results of Th. \ref{th: Very Strong Interference capacity results} to a sample channel: the InterFerence Channel with two common messages (IFC-2CM). The IFC-2CM is a classical interference channel where each transmitter is also sending a common message to the two decoders.
%
A graphical representation of this channel ran be found in Fig. \ref{fig:IFC-CM}.

We start by considering applying Cor. \ref{cor:Superposition coding achievable scheme} to this channel model.
The private message $W_{1 \sgoes 1}$ and $W_{1 \sgoes \{1,2\}}$ are both encoded by transmitter one and are thus encoded in the same codeword $U_{ 1 \sgoes \{1,2\}}^N$,since $\jv_{\rm all}=\{1,2\}$, for $R_{i \sgoes \{1,2\}}'=R_{1 \sgoes 1}+R_{1 \sgoes \{1,2\}}$. The same reasoning can be applied to the messages of the second user.
%
The achievable region with the rate-splitting in \eqref{eq:rate splitting all common} is
\pp{
& R_{i \sgoes \{1,2\} }'   & \leq I(Y_i ; X_1 , X_2 | U_{i \sgoes \{1,2\}}')  \\
& R_{1 \sgoes \{1,2\} }' + R_{2 \sgoes \{1,2\}}'  & \leq I(Y_i ; X_1 , X_2),
}
for $P_{X_i,U_{i \sgoes \{1,2\}}'}$ and with $i \in \{1,2\}$.
With the assignment in \eqref{eq:assigment Uprime to U} and \eqref{eq:realte S and Sall} we obtain the achievable region
\eas{
& R_{i \sgoes i} +  R_{i \sgoes \{1,2\} }   \leq 
I(Y_i ; X_1 , X_2 | U_{i \sgoes i}, U_{i \sgoes \{1,2\}})
\label{eq:IFC-2CM single rate 1}\\
& R_{2 \sgoes 2 } + R_{2 \sgoes \{1,2\} }    \leq 
I(Y_2 ; X_1 , X_2 | U_{2 \sgoes 2}, U_{1 \sgoes \{1,2\}})
\label{eq:IFC-2CM single rate 2}\\
& R_{1 \sgoes 1 } +  R_{1 \sgoes \{1,2\} } +   R_{2 \sgoes 2 } + R_{2 \sgoes \{1,2\} }  \leq I(Y_1 ; X_1 , X_2)
\label{eq:IFC-2CM sum rate 1} \\
& R_{1 \sgoes 1 } +  R_{1 \sgoes \{1,2\} } +   R_{2 \sgoes 2 } + R_{2 \sgoes \{1,2\} }  \leq I(Y_2 ; X_1 , X_2),
\label{eq:IFC-2CM sum rate 2}
}{\label{eq:IFC-2CM}}
%
%
%
We now match the inner bound expression in \eqref{eq:IFC-2CM} with the outer bound in Cor. \ref{cor:min cut outer bound}.
By consider $\Sv=\Sv^1=\lcb (1,1), (1,\{1,2\}) \rcb $
in Cor. \ref{cor:min cut outer bound}, we obtain the bound
$$
R_{1 \sgoes 1}+R_{1 \sgoes \{1,2\}} \leq I(Y_1; U_{1 \sgoes 1},  U_{1 \sgoes \{1,2\}} |U_{2 \sgoes 2},  U_{2 \sgoes \{1,2\}}),
$$
which is equivalent to the bound in \eqref{eq:IFC-2CM single rate 1}. The bound for \eqref{eq:IFC-2CM single rate 2} is obtained from
$\Sv=\Sv^2=\lcb (2,2), (1,\{2,2\}) \rcb$ in a similar manner.
%
%
The bound in \eqref{eq:IFC-2CM sum rate 1} cannot be matched with an outer bound from Cor. \ref{cor:min cut outer bound}, since the decoders are not required to decode all the messages.
From Cor. \ref{cor:min cut outer bound} we can obtain sum rates bounds of the form
\eas{
R_{\rm sum} & \leq I(Y_1 ; X_1, X_2| U_{1 \sgoes \{1,2\}}, U_{2 \sgoes 2},U_{1 \sgoes \{1,2\}}) \nonumber \\
            & \quad +I(Y_2;X_1 , X_2 | U_{1 \sgoes 1 } )
            \label{eq:R sum 1}\\
R_{\rm sum} & \leq I(Y_1 ; X_1, X_2| U_{1 \sgoes \{1,2\}}, U_{2 \sgoes 2}) \nonumber \\
            & \quad +I(Y_2;X_1 , X_2 | U_{1 \sgoes 1 } , ,U_{2 \sgoes \{1,2\}})
            \label{eq:R sum 2}\\
R_{\rm sum} & \leq I(Y_1 ; X_1, X_2| U_{2 \sgoes \{1,2\}}, U_{2 \sgoes 2}) \nonumber \\
            & \quad +I(Y_2;X_1 , X_2 | U_{1 \sgoes 1 } , ,U_{1 \sgoes \{1,2\}})
            \label{eq:R sum 3}\\
R_{\rm sum} & \leq I(Y_1 ; X_1, X_2| U_{2 \sgoes 2}) \nonumber \\
            & \quad +I(Y_2;X_1 , X_2 | U_{1 \sgoes 1 } , ,U_{1 \sgoes \{1,2\}},U_{2 \sgoes \{1,2\}}),
            \label{eq:R sum 4}
}
and for $R_{\rm sum}=R_{1\sgoes 1}+R_{1\sgoes \{1,2\}}+R_{2 \sgoes 2}+R_{2\sgoes \{1,2\}}$
To show that the region in \eqref{eq:IFC-2CM}, we can impose conditions as in Cor. \ref{cor:Simplified inner bound} and Cor. \ref{cor:strong inteference outer bound}.
This can be done in different ways:

\noindent
{$\bullet$\bf Remove two sum rates from the inner bound} by imposing
\pp{
\eqref{eq:IFC-2CM single rate 1}+\eqref{eq:IFC-2CM single rate 2}  \leq \max \{ \eqref{eq:IFC-2CM sum rate 1}, \eqref{eq:IFC-2CM sum rate 2} \},
}
in which case the capacity region reduces to \eqref{eq:IFC-2CM single rate 1} and \eqref{eq:IFC-2CM single rate 2}.

\noindent
{$\bullet$\bf Remove a sum rate from the inner bound and add a sum rate in the outer bound}
by setting $\eqref{eq:IFC-2CM sum rate 2} \geq \eqref{eq:IFC-2CM sum rate 1}$ and one of the following conditions
$\eqref{eq:IFC-2CM sum rate 1} \geq \eqref{eq:R sum 1}$, $\eqref{eq:IFC-2CM sum rate 1} \geq \eqref{eq:R sum 2}$, $\eqref{eq:IFC-2CM sum rate 1} \geq \eqref{eq:R sum 3}$ or $\eqref{eq:IFC-2CM sum rate 1} \geq \eqref{eq:R sum 4}$.
in which case the capacity region reduces to \eqref{eq:IFC-2CM single rate 1}, \eqref{eq:IFC-2CM single rate 2} and \eqref{eq:IFC-2CM sum rate 1}.
One can also consider the symmetric conditions obtained by setting $\eqref{eq:IFC-2CM sum rate 1} \geq \eqref{eq:IFC-2CM sum rate 2}$ and change the other conditions accordingly.

\noindent
{$\bullet$\bf Add two sum rates in the outer bounds}
by setting, for example, either $\eqref{eq:R sum 1} \geq \max \{\eqref{eq:IFC-2CM sum rate 1},\eqref{eq:IFC-2CM sum rate 2}\}$ or
$\eqref{eq:R sum 1} \geq \eqref{eq:IFC-2CM sum rate 1}, \ \eqref{eq:R sum 2} \geq \eqref{eq:IFC-2CM sum rate 2}$ or
$\eqref{eq:R sum 3} \geq \eqref{eq:IFC-2CM sum rate 1}, \ \eqref{eq:R sum 4} \geq \eqref{eq:IFC-2CM sum rate 2}$
and so on.

\section{Conclusion}

In this paper we derive the a compact representation of the capacity of a general multiple-access channel with any number of transmitters and any distribution of messages among the transmitters.
From this result we derive a capacity result in a certain class of a general class of channels with any number of transmitters, receivers, and any distribution of messages.
In this class of channels, that we denote as in ``very strong interference'', there is no rate loss in having every decoder decode all the messages and the capacity region reduces to the intersection of the multiple-access channels from all the encoders to each decoder.
To exemplify this result we derive capacity in the ``very strong interference'' regime for the interference channel where each decoder is sending a message to the intended receiver and also a message to both receivers.

\section*{Acknowledgment}
The author would like to thank Prof. Gerhard Kramer  for the insightful
discussions and helpful comments.

\bibliographystyle{IEEEtran}
\bibliography{steBib1}

\end{document}